\documentclass[3p]{elsarticle}
\usepackage{amssymb}
\usepackage{amsmath}
\usepackage{epsfig}
\usepackage{graphicx}
\usepackage{mathrsfs}
\usepackage{algorithmic}
\usepackage{algorithm}
\newtheorem{example}{Example}
\def\diag{\mathop{\rm diag}\nolimits}

\renewcommand{\dj}{d\kern-0.4em\char"16\kern-0.1em}
\renewcommand{\DJ}{\raise0.3ex\hbox{-}\kern-0.36em D}
\journal{Journal of Computational Physics}

\begin{document}

\begin{frontmatter}
\title{Piecewise linear transformation in diffusive flux discretization}

\author[cerni]{D. Vidovi\'{c}\corref{cor1}}
\ead{dragan.vidovic.jcerni@gmail.com}
\author[cerni]{M. Dotli\'{c}}
\ead{milan.dotlic@jcerni.co.rs}
\author[rudarski]{M. Pu\v si\'c}
\ead{mpusic@ptt.rs}
\author[cerni]{B. Pokorni}
\ead{bpokorni.jci@gmail.com}

\address[cerni]{Jaroslav \v Cerni Institute, Jaroslava \v Cernog 80, 11226 Pinosava, Belgrade, Serbia}
\address[rudarski]{University of Belgrade, Faculty of Mining and Geology, \DJ u\v sina 7, 11000 Belgrade, Serbia}
\cortext[cor1]{Corresponding author. Tel. +381 64 331 5976; fax: +381 11 390 6480}

%\renewcommand{\thefootnote}{\fnsymbol{footnote}}
%\footnotetext[2]{Jaroslav \v Cerni Institute, Jaroslava \v Cernog 80, 11226 Pinosava, Belgrade, Serbia}
%\footnotetext[3]{draganvid@gmail.com}
%\footnotetext[4]{milandotlic@gmail.com}
%\footnotetext[6]{University of Belgrade, Faculty of Mining and Geology, \DJ u\v sina 7, 11000 Belgrade, Serbia (mpusic@ptt.rs)}
%\footnotetext[5]{bpokorni.jci@gmail.com}

%\renewcommand{\thefootnote}{\arabic{footnote}}

\begin{abstract}
To ensure the discrete maximum principle or solution positivity in finite volume schemes, diffusive flux is sometimes discretized as a conical combination of finite differences. Such a combination may be impossible to construct along material discontinuities using only cell concentration values. This is often resolved by introducing auxiliary node, edge, or face concentration values that are explicitly interpolated from the surrounding cell concentrations. We propose to discretize the diffusive flux after applying a local piecewise linear coordinate transformation that effectively removes the discontinuities. The resulting scheme does not need any auxiliary concentrations and is therefore remarkably simpler, while being second-order accurate under the assumption that the structure of the domain is locally layered.
\end{abstract}


%\begin{AMS}
%35B50, 35J25, 65D05, 65D18, 65N08, 65N22
%\end{AMS}

%\pagestyle{myheadings}
%\thispagestyle{plain}
%\markboth{D. VIDOVI\'C, M. DOTLI\'C, M. PU\v SI\'C AND B. POKORNI}{PIECEWISE LINEAR TRANSFORMATION}

\end{frontmatter}

\section{Introduction}

Diffusion in an anisotropic discontinuous environment plays a role in various fields of engineering, such as subsurface flows. Steady state diffusion of a solute with concentration $C$ in a bounded domain $\Omega\subset\mathbb R^3$ is modeled by the following boundary problem:
\begin{eqnarray}
\nabla\cdot\mathbf u=g,\label{eq} \\
\mathbf u=-\mathbb D\nabla C, \\
C=g_\mathrm D \quad \text{on } \Gamma_\mathrm D, \label{dirichlet} \\
\mathbf u\cdot\mathbf n=g_\mathrm N \quad \text{on } \Gamma_\mathrm N, \\
\mathbf u\cdot\mathbf n = \Psi(C-g_\mathrm R) \quad \text{on } \Gamma_\mathrm R, \label{robin}
\end{eqnarray}
where $\mathbf u$ is the velocity, $g$ is the volumetric source term, $\mathbf n$ is the unit vector normal to $\partial\Omega$ pointing outward, $\Psi$ is the transfer coefficient, $\Gamma_\mathrm D\cup\Gamma_\mathrm R=\overline{\Gamma_\mathrm D\cup\Gamma_\mathrm R}$, $\Gamma_\mathrm D\cup\Gamma_\mathrm N\cup\Gamma_\mathrm R=\partial\Omega$, $\Gamma_\mathrm D\cup\Gamma_\mathrm R\neq\emptyset$, and $\Gamma_\mathrm D$, $\Gamma_\mathrm N$, and $\Gamma_\mathrm R$ are mutually disjoint. Diffusion tensor $\mathbb D$ is symmetric, positive definite, and piecewise continuous. Connected subsets of $\Omega$ in which $\mathbb D$ is continuous are called {\em material zones}, and the interfaces between them are referred to as {\em material interfaces}. Presumably, mesh faces coincide with material interfaces, i.e. $\mathbb D$ is continuous within mesh cells.

Various numerical schemes are used to solve this problem. Most of them produce non-physical oscillations and negative concentration values in particular cases. Nevertheless, certain schemes are specifically designed to address these issues. One such finite volume scheme appeared in \cite{Lep05} and was further developed in \cite{Dan09,Lip07,Lip09,Vas08,Vid11,Vid13,Yua08,Lip13}. These schemes do not satisfy the maximum principle, but they guarantee that the concentration does not become negative.

In addition to the primary concentration unknowns associated with mesh cells, these schemes use auxiliary concentration values located in faces and elsewhere. Face concentration values are easily determined from the continuity, but some kind of interpolation must be used to determine other auxiliary values. The interpolation method presented in \cite{Vid13} performs this task using piecewise linear interpolation and convex combinations. It satisfies the maximum principle and is second order accurate even when it uses interpolation nodes at the opposite sides of a material discontinuity. Since a simple brute force search for collocation points that form a convex combination can result in a combinatorial explosion, in \cite{Vid13} we proposed a complex but efficient alternative search algorithm based on Delaunay triangulations.

In this paper we deploy the piecewise linear transformation introduced in \cite{Vid13} directly in the velocity decomposition, without any auxiliary concentration variables. Such a scheme is simpler than \cite{Vid13} because the complicated construction of convex combinations is avoided. In addition, fluxes over discontinuities do not need special treatment, resulting in further simplification.

The paper is organized as follows: in \S \ref{sec:fluxes} we explain how to use the piecewise linear transformation to obtain a one-side flux approximation. In \S \ref{sec:scheme} the one-side fluxes are combined in the usual fashion to obtain a two-point scheme. Euler implicit temporal discretization of the time-dependent problem is presented in \S \ref{sec:transient}. Numerical tests presented in \S \ref{sec:examples} show that the accuracy and the convergence rate do not change much in comparison to \cite{Vid13}, and thus the simpler scheme should be the method of choice. Moreover, it is shown that a single iteration is sufficient to maintain the second order accuracy in a time-dependent problem.

%\section{Approximation of boundary fluxes}

%If $f$ is a boundary face belonging to cell $c$, with $\mathbf x_f$ we denote the projection of $\mathbf x_c$ in the direction of vector $\mathbb D_f\mathbf n_f$ on the plane containing face $f$
%\begin{equation}
%\mathbf x_f = \mathbf x_c + \frac{(\overline{\mathbf x}_f-\mathbf x_c)\cdot\mathbf n_f}{\mathbf n^T\mathbb D\mathbf n_f}\mathbb D\mathbf n_f,\label{projection}
%\end{equation}
%where diffusion tensor $\mathbb D_f$ is evaluated at the centroid of face $f$, and $\overline{\mathbf x}_f$ is any point belonging to face $f$.
%If $f\subseteq\Gamma_\mathbb D$ then we take
%\begin{equation}
%u_f\approx-|f||\boldsymbol\ell_f|\frac{g_\mathbb D(\mathbf x_f)-C_c}{|\mathbf x_f-\mathbf x_c|}.\label{DirichletFlux}
%\end{equation}
%Otherwise, if $f\subseteq\Gamma_R$ then
%\begin{equation}
%u_f\approx-|f||\boldsymbol\ell_f|\frac{C_f-C_c}{|\mathbf x_f-\mathbf x_c|} \quad \text{ and } \quad u_f=|f|\Psi(C_f-g_R(\mathbf x_f)).\label{RobinFlux}
%\end{equation}
%Eliminating the unknown face concentration $C_f$ gives
%\begin{equation}
%u_f\approx |f|\frac{\Psi|\boldsymbol\ell_f|/|\mathbf x_f-\mathbf x_c|}{\Psi+|\boldsymbol\ell_f|/|\mathbf x_f-\mathbf x_c|}(C_c-g_R(\mathbf x_f)).
%\end{equation}

\section{Fluxes and piecewise linear transformation}\label{sec:fluxes}

The piecewise linear transformation used here to approximate the flux was constructed in \cite{Vid13}. The main steps are repeated for convenience.

\begin{figure}[htbp]
   \begin{center}
      \includegraphics[height=50mm]{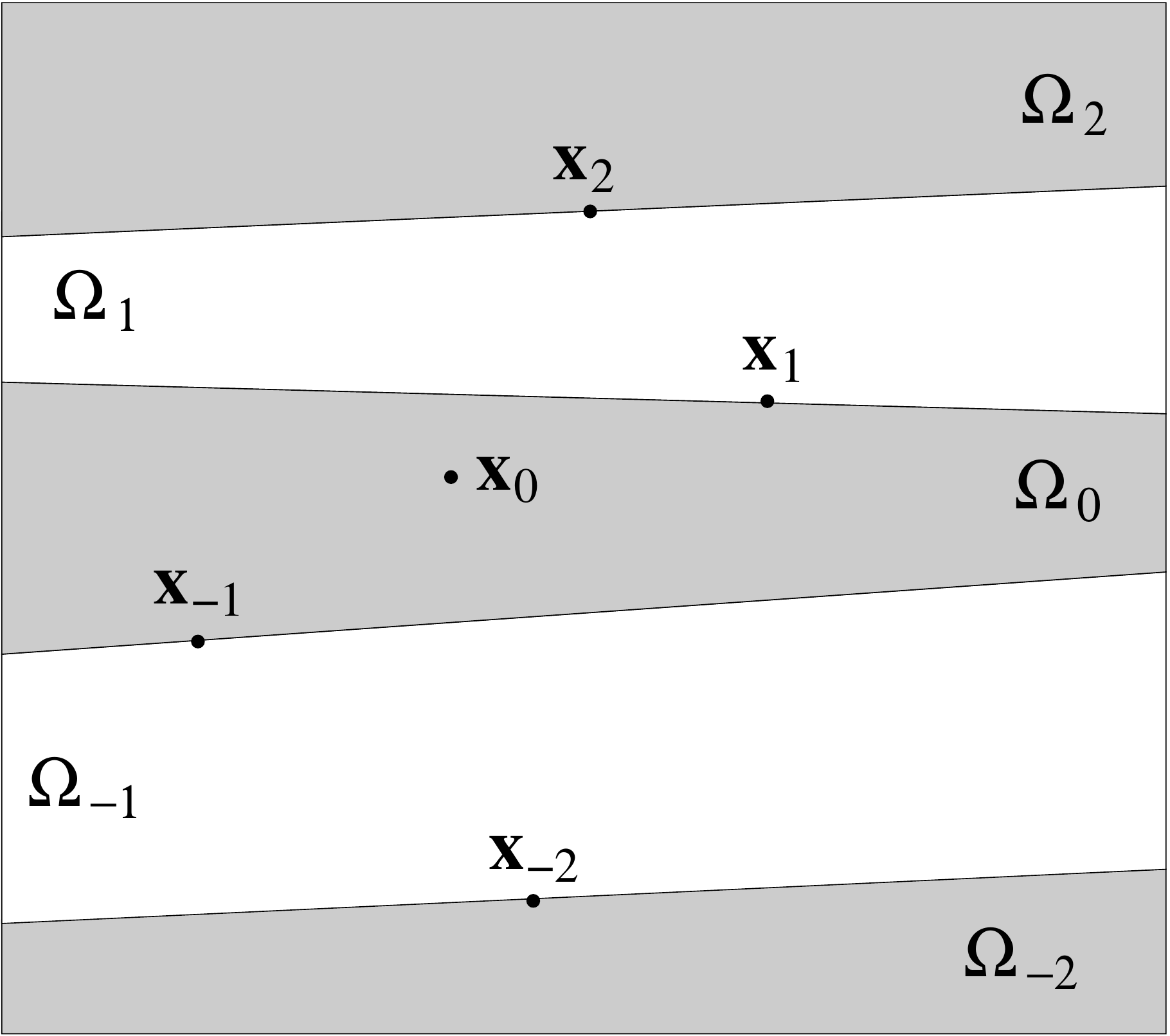}
      \caption{It is assumed that the structure of the domain is locally layered.}
      \label{fig:stack}
   \end{center}
\end{figure}
We assume that some neighborhood of point $\mathbf x_0\in\Omega$ consists of layers $\Omega_{-m}$, $\dots$, $\Omega_n$, $m\geq0$, $n\geq0$, with interfaces that are either planes or approximated by planes (see Fig. \ref{fig:stack}). The diffusion tensor is allowed to have discontinuities between layers, but it is assumed to be constant or almost constant within each layer, thus $\mathbb D|_{\Omega_i}=\mathbb D_i$. This assumption needs to hold only locally --- the diffusion tensor is allowed to vary smoothly within layers on a larger scale. We assume that the concentration changes linearly in each layer
\begin{equation}
C(\mathbf x)=C_i+\mathbf G_i\cdot(\mathbf x-\mathbf x_i), \qquad \mathbf x_i\in\overline{\Omega}_i. \label{piecewise}
\end{equation}
This function must satisfy two conditions:
\begin{enumerate}
\item It must be continuous, and
\item The flux through each interface must be continuous.
\end{enumerate}
The two conditions eliminate all but four degrees of freedom in the piecewise linear function $C(\mathbf x)$, which may be written as
\begin{equation}
C(\mathbf x)=C_0+\mathbf G_0\cdot F(\mathbf x),\label{trans}
\end{equation}
where $F:\mathbb R^3\rightarrow\mathbb R^3$ is a piecewise linear transformation depending on the diffusion tensors and geometry but not on the concentration. In \cite{Vid13} we derived explicit formulas to compute $F$.

In \cite{Vid13} we evaluate (\ref{trans}) to obtain the concentration values at points. In contrast, in this paper (\ref{trans}) is not evaluated at points: instead, the gradient of (\ref{trans}) is used to approximate the normal velocity component $\mathbf n_f\cdot\mathbf u$ in some mesh face $f$. We assume that this face belongs to cell $T$ and that $\mathbf n_f$ points outside from this cell. For point $\mathbf x_0$ in (\ref{piecewise}) we take the centroid $\mathbf x_T$; thus $C_0=C_T$ in (\ref{trans}). We need to determine the three components of $\mathbf G_0$. These are found by solving a linear system consisting of three equations of the following types:
\begin{equation}
F(\mathbf x_{\mathcal{T}})\cdot\mathbf G_0 =C_{\mathcal{T}}-C_T \quad \text{if } \mathcal{T} \text{ is a mesh cell,}\label{dirichleteq}
\end{equation}
\begin{equation}
F(\mathbf x_n)\cdot\mathbf G_0=g_\mathrm D(\mathbf x_n)-C_T \quad \text{if } \mathbf x_n \text{ is a node in } \Gamma_\mathrm D,\label{dirichletnodeeq}
\end{equation}
\begin{equation}
F(\mathbf x_f)\cdot\mathbf G_0=g_\mathrm D(\mathbf x_f)-C_T \quad \text{if } f \text{ is a face in } \Gamma_\mathrm D,\label{dirichletfeq}
\end{equation}
\begin{equation}
\mathbf n_f^\mathrm T\mathbb D(\mathbf x_f)\nabla F(\mathbf x_f)\mathbf G_0=-g_\mathrm N(\mathbf x_f) \quad \text{if } f \text{ is a face in } \Gamma_\mathrm N,
\end{equation}
\begin{equation}
\left(F(\mathbf x_f)+\frac1\Psi\mathbf n_f^\mathrm T\mathbb D(\mathbf x_f)\nabla F(\mathbf x_f)\right)\cdot\mathbf G_0=g_\mathrm R(\mathbf x_f)-C_T \quad \text{if } f \text{ is a face in } \Gamma_\mathrm R.\label{robineq}
\end{equation}
Point $\mathbf x_f$ is the centroid of face $f$. The last equation follows from
\begin{equation}
-\mathbf n_f^\mathrm T\mathbb D(\mathbf x_f)\nabla F(\mathbf x_f)\mathbf G_0=\Psi(C_T+\mathbf G_0\cdot F(\mathbf x_f)-g_\mathrm R(\mathbf x_f))
\end{equation}
and may be written in a form similar to (\ref{dirichletnodeeq}) and (\ref{dirichletfeq}):
\begin{equation}
F\left(\mathbf x_f+\frac1\Psi \mathbb D(\mathbf x_f)\mathbf n_f\right)\cdot\mathbf G_0=g_\mathrm R(\mathbf x_f)-C_T \quad \text{if } f \text{ is a face in } \Gamma_\mathrm R.\label{robineq1}
\end{equation}
%Collocation point $\mathbf x_f$ associated with face $f$ where Dirichlet or Robin condition is specified is determined as the projection of the adjecent cell's collocation pojnt $\mathbf x_c$ in the direction of vector $\mathbb D_f\mathbf n_f$ on the plane containing face $f$
%\begin{equation}
%\mathbf x_f = \mathbf x_c + \frac{(\overline{\mathbf x}_f-\mathbf x_c)\cdot\mathbf n_f}{\mathbf n^T\mathbb D\mathbf n_f}\mathbb D\mathbf n_f \label{projection}
%\end{equation}
%if this point belongs to face $f$. Otherwise, for $\mathbf x_f$ we take the intersection of face $f$ boundary and the segment connecting projection (\ref{projection}) with the centroid of face $f$. If $f\in\Gamma_N$ then for $\mathbf x_f$ we take the face centroid.

Three equations of the form (\ref{dirichleteq})--(\ref{robineq}) constitute a linear system
\begin{equation}
M\mathbf G_0=\mathbf r.\label{system}
\end{equation}
If this system has a unique solution $\mathbf G_0=M^{-1}\mathbf r$, then the normal velocity component at face $f$ can be approximated as
\begin{equation}
\mathbf n_f\cdot\mathbf u\approx-\mathbf n_f^\mathrm T\mathbb D(\mathbf x_f)M^{-1}\mathbf r=-\boldsymbol\alpha\cdot\mathbf r\label{uapprox}
\end{equation}
because $f\subset\overline\Omega_0$ and $\nabla F\equiv I$ in $\Omega_0$. Coefficient vector $\boldsymbol\alpha$ is determined by solving
\begin{equation}
M^\mathrm T\boldsymbol\alpha=\mathbb D(\mathbf x_f)\mathbf n_f.
\end{equation}
Equations that make up the system (\ref{system}) are chosen in such a way that the obtained coefficients $\boldsymbol\alpha$ are non-negative. The search for these equations is performed by testing all combinations of neighboring cells and boundary conditions belonging to the {\em candidate set}, in the same way that the brute force strategy in \cite{Vid13} works. The candidate set is initialized with cell $T$, cells that share a face with $T$, boundary faces of $T$ and Dirichlet boundary nodes of $T$. If equations leading to non-negative $\boldsymbol\alpha$ are not found, the candidate set is extended by adding all neighboring cells, boundary faces and Dirichlet boundary nodes of the current member cells, until non-negative $\boldsymbol\alpha$ is obtained.

If matrix $M$ in (\ref{system}) is ill-conditioned, this typically results in large coefficients $\boldsymbol\alpha$. The resulting interpolation is inaccurate, so the combination is discarded if some $\alpha_i\|\mathbf t_i\|/\|\mathbb D\mathbf n_f\|$ is larger than a suitably chosen value $A_\text{max}$ discussed in the next section, where $\mathbf t_i$ is the vector multiplying $\mathbf G_0$ in the corresponding equation (\ref{dirichleteq})--(\ref{robineq}).

If more than one conical combination is found, we apply a strategy similar to the one used in \cite{Dan09}: starting with combinations containing the cell that shares the face $f$ with the cell $T$, we choose the first encountered conical combination such that $\max_i(\alpha_i m_i)/\|\mathbb D(\mathbf x_f)\mathbf n_f\|_2\leq1$, where $m_i$ is the 2-norm of the $i^{\text{th}}$ row of matrix $M$. If no such combination is encountered, then we choose the one for which this ratio is minimal.

\subsection{Alternative: harmonic averaging points}

Let points $\mathbf x_1$ and $\mathbf x_2$ be separated by a material interface. The same two conditions leading to transformation $F$ were used in \cite{Age09} to compute the concentration at a suitably chosen point at the interface as a convex combination of concentration values at $\mathbf x_1$ and $\mathbf x_2$. Such so-called {\em harmonic averaging point} can be used in (\ref{dirichleteq}) in the place of $\mathcal{T}$, as an alternative to transformation $F$. The two approaches are equivalent under certain conditions:
\begin{itemize}
\item Either $\mathbf x_1$ or $\mathbf x_2$ is equal to $\mathbf x_T$;
\item The concentration at the harmonic averaging point is evaluated implicitly;
\item Points $\mathbf x_1$ and $\mathbf x_2$ are in neighboring layers;
\item Neumann and Robin boundary conditions in faces not belonging to $\overline\Omega_0$ are not used in (\ref{system}).
\end{itemize}
The notion of harmonic averaging points could be extended to interpolate over multiple layers and to deploy boundary conditions in a similar way this is done here.

\section{Non-linear two-point scheme}\label{sec:scheme}

As usual, finite volume discretization is performed applying the divergence theorem to the integral of (\ref{eq}) over the mesh cell $T$:
\begin{equation}
\oint_T \mathbf u\cdot\mathbf n{\rm d}S=\sum_f\chi_{T,f}u_f=\int_Tg{\rm d}T, \qquad u_f=\int_f\mathbf u\cdot\mathbf n{\rm d}S,\label{fv}
\end{equation}
where $\mathbf n_f$ is the fixed unit normal vector associated with face $f$, and $\chi_{T,f}=1$ if $\mathbf n_f$ points outside of $T$, or $\chi_{T,f}=-1$ otherwise.

The two-point scheme is constructed in a similar way as in \cite{Dan09,Vid11}, but simpler because the discontinuities do not need to be treated separately. Approximation (\ref{uapprox}) can be written as
\begin{equation}
\mathbf n_f\cdot\mathbf u\approx\sum_i\alpha_i(C_T-C_i)+\sum_j\alpha_jg_\mathrm N(\mathbf x_j).\label{uapprox1}
\end{equation}
The convex combination of such expressions that correspond to the cells $T^+$ and $T^-$ sharing the internal face $f$
\begin{equation}
\frac{u_f}{|f|}\approx\mu_+\left(\sum_i\alpha_i^+(C_+-C^+_i)+\sum_j\alpha_j^+g_\mathrm N(\mathbf x_j^+)\right)-\mu_-\left(\sum_k\alpha_k^-(C_--C^-_i)+\sum_l\alpha_l^-g_\mathrm N(\mathbf x_l^-)\right)\label{theflux}
\end{equation}
is chosen in such a way that the contributions of concentration values other than $C_+$ and $C_-$ cancel, together with the contributions of inflow Neumann faces. Here $C_\pm$ denotes the concentration in cell $T^\pm$. Thus weights $\mu_+$ and $\mu_-$ are determined such that
\begin{equation}
\mu_++\mu_-=1, \qquad -\mu_+d_++\mu_-d_-=0,\label{mus}
\end{equation}
where
\begin{equation}
d_\pm=\sum_{\substack{i \\ \mathbf x_i\neq\mathbf x_\mp}}\alpha_iC_i^\pm-\sum_{\substack{j \\ g_\mathrm N(\mathbf x_j^\pm)<0}}\alpha_j^\pm g_\mathrm N(\mathbf x_j^\pm).\label{des}
\end{equation}
If $d_++d_-\neq0$, then
\begin{equation}
\mu_+=\frac{d_-}{d_++d_-}, \qquad \mu_-=\frac{d_+}{d_++d_-}.\label{mu}
\end{equation}
Otherwise we take $\mu_\pm=\tfrac12$.

In this way flux approximation (\ref{theflux}) reduces to a two-point formula
\begin{equation}
u_f\approx M_f^+C_+-M_f^-C_-+r_f,\label{twopoint}
\end{equation}
where
\begin{equation}
M_f^\pm=|f|\left(\mu_\pm\sum_i\alpha_i^\pm+\mu_\mp\sum_{\substack{k \\ \mathbf x_k=\mathbf x_\pm}}\alpha_k^\mp\right), \qquad r_f=|f|\left(\mu_+\sum_j\alpha_j^+g_\mathrm N(\mathbf x_j^+)-\mu_-\sum_l\alpha_l^-g_\mathrm N(\mathbf x_l^-)\right).\label{mdef}
\end{equation}
Fluxes through Dirichlet faces are discretized in the same way as in \cite{Dan09,Vid11}, treating the boundary face as a flat cell of zero volume.  The same technique is applied to Robin faces by treating the Robin condition at $\mathbf x_f$ as a Dirichlet condition specified at $\mathbf x_f+\tfrac1\Psi \mathbb D(\mathbf x_f)\mathbf n_f$ (see (\ref{robineq1})).

Experiments have shown that it is better to accept relatively large coefficients $\boldsymbol\alpha$ than to extend the candidate set, which also increases the error due to the increased distances between points. From (\ref{des}) and (\ref{mu}) we see that large coefficients in a one-side flux cause the weight assigned to this flux to be small and the weight assigned to the opposite flux to be large. In this way the influence of inaccurate interpolation vanishes. Owing to this fortunate circumstance, we only discard combinations if $\alpha_i\|\mathbf t_i\|/\|\mathbb D\mathbf n_f\|$ is larger than $A_\text{max}=1000$ for some $i$.

Finite volume discretization with fluxes as in (\ref{twopoint}) results in a non-linear system $A(\mathbf C)\mathbf C=\mathbf b(\mathbf C)$, where vector $\mathbf C$ contains concentration values in all cells.

Starting with some $\mathbf C^0$, consecutive iterations are computed by solving the linearized system $A(\mathbf C^n)\mathbf C^{n+1}=\mathbf b(\mathbf C^n)$ until the convergence criterion
\begin{equation}
\frac{\|A(\mathbf C^n)\mathbf C^n-\mathbf b(\mathbf C^n)\|}{\|\mathbf b(\mathbf C^n)\|}<\varepsilon\label{convergencecriterion}
\end{equation}
is met.

If $\mathbf C^n$ is non-negative, then $A(\mathbf C^n)$ is an M-matrix because coefficients $M^\pm_f$ in (\ref{twopoint}) are non-negative, which follows from the non-negativity of $d_\pm$, (\ref{mu}), and (\ref{mdef}). In the absence of Neumann outflow boundaries and if $g_\mathrm D\geq0$ and $g_\mathrm R\geq0$, the right-hand side is non-negative. Therefore $\mathbf C^{n+1}$ is non-negative as well, and by induction the concentration is non-negative in each iteration.

\section{Time-dependent problem}\label{sec:transient}

A transient problem is obtained by adding a time derivative to the continuity equation (\ref{eq}):
\begin{equation}
\frac{\partial C}{\partial t}+\nabla\cdot\mathbf u=g
\end{equation}
when $t>t_0$. For this equation to have a unique solution, in addition to boundary conditions (\ref{dirichlet})-(\ref{robin}), one also needs to specify an initial condition $C|_{t=t_0}=C_0$. Condition $\Gamma_\mathrm D\cup\Gamma_\mathrm R\neq\emptyset$ is no longer necessary.

Using Euler implicit discretization, we obtain
\begin{equation}
|T|\frac{C_T^{n+1}-C_T^n}{\Delta t^{n+1}}+\sum_f\chi_{T,f}\left(M_f^{+,n+1}C_+^{n+1}- M_f^{-,n+1}C_-^{n+1}+r_f^{n+1}\right)=\int_Tg^{n+1}{\rm d}T\label{nonlintransient}
\end{equation}
in the place of (\ref{fv}), where $\Delta t$ is the time step and the last superscript indicates the time level.

At each time level this non-linear system is solved using Picard linearization
\begin{equation}
|T|\frac{C_T^{n+1,m+1}-C_T^n}{\Delta t^{n+1}}+\sum_f\chi_{T,f}\left(M_f^{+,n+1,m}C_+^{n+1,m+1}- M_f^{-,n+1,m}C_-^{n+1,m+1}+r_f^{n+1,m}\right) =\int_Tg^{n+1,m}{\rm d}T, \label{picardtransient}
\end{equation}
where the superscript after the time level indicates the iteration number.

To preserve the second-order accuracy, one must take $\Delta t=\mathcal O(h^2)$, where mesh parameter $h$ is proportional to the longest edge length. 

%The error introduced at each time step by Euler implicit time discretization is $\mathcal O(\Delta t^2)$ because
%\begin{equation}
%C_T^{n+1}=C_T^n+\Delta t^{n+1}\dot C_T^{n+1}, \qquad \dot C_T^{n+1} = \frac{C_T^{n+1}-C_T^n}{\Delta t^{n+1}},
%\end{equation}
%and $\dot C_T^n$ is a first-order accurate approximation of the time derivative at $t^{n+1}$. This error accumulates to $\mathcal O(\Delta t)$ at the end of the time stepping process.

Inspired by \cite{Kav01,Kav02a,Kav02b}, for the initial approximation we take
\begin{equation}
C_T^{n+1,0}=\max (C_T^n+\Delta t^{n+1} \dot C_T^n, 0), \qquad \dot C_T^n = \frac{C_T^n-C_T^{n-1}}{\Delta t^n}.\label{extrapolation}
\end{equation}
Such an initial value violates the mass conservation property, but it is used only to compute coefficients $M_f^\pm$ in the first iteration. The conservation and positivity of subsequent iterations follow from the matrix structure. Moreover, since the error of the initial value is already $\mathcal O(\Delta t^2)$, a single iteration in each time step is enough to maintain the $\mathcal O(\Delta t)=\mathcal O(h^2)$ accuracy at the end of the time stepping process.

Extrapolation (\ref{extrapolation}) cannot be used in the first time step because $\dot C_T^0$ is not known. It is suggested in \cite{Kav01,Kav02a,Kav02b} to calculate $\dot C_T^0$ from the truncation error obtained by substituting the initial solution into the stationary problem. However, finite volume schemes are inconsistent in the finite difference sense so $\dot C_T^0$ obtained in this way is inaccurate. One can take $C^{1,0}_T=C^0_T$ instead. If a single iteration is performed, the error $\mathcal O(\Delta t)$ is introduced only at the first time step while all other steps add $\mathcal O(\Delta t^2)$, so the total error is still $\mathcal O(\Delta t)$. Alternatively, one may compute a few iterations in the first time step.

\section{Examples}\label{sec:examples}

The following examples demonstrate that, on average, the proposed scheme is not less accurate and does not require more iterations than the more complicated version presented in \cite{Vid13}.

We report the numbers of iterations $n_6$ and $n_9$ until convergence is achieved with $\varepsilon=10^{-6}$ and $\varepsilon=10^{-9}$, respectively.  %Many applications require lower accuracy, but if an automatic parameter estimation method deploying a numerical Jacobian is to be used then a higher accuracy is necessary.
Linear systems are solved using the BiCGStab solver with an accuracy of $10^{-12}$.

The following norms are used to quantify the concentration and flux errors:
\begin{equation}
\varepsilon^C_2=\left[\frac{\sum_T(C(\mathbf x_T)-C_T)^2|T|}{\sum_T(C(\mathbf x_T))^2|T|}\right]^{1/2}, \quad \varepsilon_\text{max}^C=\frac{\max_T|C(\mathbf x_T)-C_T|}{\left[\sum_T(C(\mathbf x_T))^2|T|/\sum_T|T|\right]^{1/2}},\nonumber
\end{equation}
\begin{equation}
\varepsilon^{\mathbf u}_2 = \left[\frac{\sum_f(\mathbf u(\mathbf x_f)-u_f/|f|)^2|f|}{\sum_f(\mathbf u(\mathbf x_f)\cdot\mathbf n_f)^2|f|}\right]^{1/2}, \quad \varepsilon^{\mathbf u}_\text{max}=\frac{\max_f|\mathbf u(\mathbf x_f)\cdot\mathbf n_f-u_f/|f||}{\left[\sum_f(\mathbf u(\mathbf x_f)\cdot\mathbf n_f)^2|f|/\sum_f|f|\right]^{1/2}},
\end{equation}
where $C_T$ is the computed concentration in cell $T$, $u_f$ is the computed flux through face $f$, and $C(\mathbf x)$ and $\mathbf u(\mathbf x)$ are the exact concentration and velocity functions, respectively.

\begin{example}\label{ex:discontiniousdanilov}

This is a modified version of an example that appeared in \cite{Dan09}. The domain is the unit cube. The diffusion tensor is
\begin{equation}
\mathbb D=\left\{\begin{array}{c}\mathbb D_1 \quad \text {if } x<0.5 \\ \mathbb D_2 \quad \text {otherwise}\end{array}\right., \quad \mathbb D_1=\left[\begin{array}{c c c}3&1&0\\1&3&0\\0&0&1\end{array}\right], \quad \mathbb D_2=\left[\begin{array}{c c c}10&3&0\\3&1&0\\0&0&1\end{array}\right].
\end{equation}
The source term is chosen such that the exact solution is 
\begin{equation}
C=\left\{\begin{array}{c}1-2y^2+4xy+2y+6x \quad \text{if } x<0.5,\\3.5-2y^2+2xy+x+3y \quad \text{otherwise.}\end{array}\right.\label{danilovSolution}
\end{equation}
The exact concentration is set at $z=0$ and $z=1$, the exact flux is specified at $y=0$ and $y=1$, and the Robin condition is prescribed at $x=0$ and $x=1$. We take $g_\mathrm R|_{x=0}\equiv0$ and $g_\mathrm R|_{x=1}\equiv10$, while $\Psi$ is chosen so (\ref{danilovSolution}) is the solution.

This problem was solved on three types of meshes: cubic, triangular prismatic, and tetrahedral%, with mesh parameter $h$ proportional to the longest edge length
. The errors and iteration numbers given in Table \ref{disDanErr} show that there is little difference in the accuracy and convergence rate, compared with the previous version of the scheme \cite{Vid13}.
\begin{table}[hbtp]
   \footnotesize
   \begin{center}
      \begin{tabular}{|c|c c c c|c c c c|}
         \hline
         & \multicolumn{4}{|c|}{Current scheme} & \multicolumn{4}{|c|}{Previous scheme} \\
         \hline
         $h$ & 1/10 & 1/20 & 1/40 & 1/80 & 1/10 & 1/20 & 1/40 & 1/80 \\
         \hline
         \multicolumn{9}{|c|}{Hexahedral grids} \\
         \hline
         $\varepsilon_2^c$ & 7.64e-4 & 1.98e-4 & 5.09e-5 & 1.31e-5 & 7.15e-4 & 1.91e-4 & 4.97e-5 & 1.25e-5 \\
         $\varepsilon_\text{max}^c$ & 3.12e-3 & 9.88e-4 & 3.08e-4 & 9.43e-5 & 3.00e-3 & 1.44e-3 & 5.90e-4 & 2.19e-4 \\
         $\varepsilon_2^{\mathbf u}$ & 4.53e-3 & 1.89e-3 & 7.40e-4 & 2.76e-4 & 3.89e-3 & 1.48e-3 & 5.46e-4 & 1.98e-4 \\
         $\varepsilon_\text{max}^{\mathbf u}$ & 2.88e-2 & 1.71e-2 & 9.18e-3 & 4.61e-3 & 1.85e-2 & 9.94e-3 & 5.07e-3 & 2.60e-3 \\
         $n_6$ & 28 & 52 & 81 & 107 & 38 & 62 & 87 & 115 \\
         $n_9$ & 45 & 84 & 143 & 205 & 61 & 100 & 154 & 220 \\
         \hline
         \multicolumn{9}{|c|}{Prismatic grids} \\
         \hline
         $\varepsilon_2^c$ & 4.78e-4 & 1.05e-4 & 2.48e-5 & 5.64e-6 & 4.41e-4 & 9.74e-5 & 2.31e-5 & 5.49e-6 \\
         $\varepsilon_\text{max}^c$ & 4.87e-3 & 2.25e-3 & 6.68e-4 & 2.95e-4 & 3.92e-3 & 1.68e-3 & 5.27e-4 & 2.17e-4 \\
         $\varepsilon_2^{\mathbf u}$ & 4.04e-3 & 2.21e-3 & 1.15e-3 & 5.84e-6 & 3.92e-3 & 2.16e-3 & 1.14e-3 & 5.83e-4 \\
         $\varepsilon_\text{max}^{\mathbf u}$ & 4.31e-2 & 2.35e-2 & 1.38e-2 & 7.74e-3 & 2.81e-2 & 1.55e-2 & 8.10e-3 & 4.50e-3 \\
         $n_6$ & 39 & 65 & 88 & 110 & 50 & 72 & 95 & 117 \\
         $n_9$ & 63 & 112 & 164 & 230 & 82 & 126 & 179 & 240 \\
         \hline
         \multicolumn{9}{|c|}{Tetrahedral grids} \\
         \hline
         $\varepsilon_2^c$ & 5.70e-4 & 1.41e-4 & 2.94e-5 & 7.45e-6 & 6.83e-4 & 1.65e-4 & 3.20e-5 & 7.42e-6 \\
         $\varepsilon_\text{max}^c$ & 6.15e-3 & 2.32e-3 & 7.59e-4 & 2.61e-4 & 8.96e-3 & 4.30e-3 & 1.02e-3 & 3.68e-4 \\
         $\varepsilon_2^{\mathbf u}$ & 5.15e-3 & 2.37e-3 & 1.15e-3 & 5.59e-4 & 5.06e-3 & 2.37e-3 & 1.15e-3 & 5.58e-4 \\
         $\varepsilon_\text{max}^{\mathbf u}$ & 3.47e-2 & 1.49e-2 & 8.18e-3 & 2.61e-3 & 3.15e-2 & 2.09e-2 & 8.17e-3 & 6.03e-3 \\
         $n_6$ & 96 & 159 & 220 & 279 & 97 & 158 & 226 & 290 \\
         $n_9$ & 147 & 260 & 384 & 534 & 150 & 259 & 393 & 549 \\
         \hline
      \end{tabular}
   \end{center}
   \caption{Errors and iteration numbers in Example \ref{ex:discontiniousdanilov}.}
   \label{disDanErr}
\end{table}
\end{example}

\begin{example}\label{ex:sinuslepotier}

This is a modified version of an example used in \cite{Vid13}. The domain is again a unit cube. The diffusion tensor is
\begin{equation}
\mathbb D=(1+0.25\cos (x+y-z))\cdot\left\{\begin{array}{c}\mathbb D_1 \quad \text{if } x<0.5 \\ \mathbb D_2 \quad \text{otherwise,}\end{array}\right. 
\end{equation}
\begin{equation}
\mathbb D_1=\left[\begin{array}{c c c}1&0&0\\0&1&0\\0&0&1\end{array}\right], \quad \mathbb D_2=\left[\begin{array}{c c c}100&0&0\\0&0.01&0\\0&0&1\end{array}\right].
\end{equation}
The source term is chosen such that the exact solution is
\begin{equation}
C=\left\{\begin{array}{c}\cos(\pi x)\sin(\pi y)+1 \quad \text{if } x<0.5,\\0.01 \cos(\pi x)\sin(\pi y)+1 \quad \text{otherwise.}\end{array}\right.
\end{equation}
The exact concentration is prescribed at $x=1$, $y=0$, and $y=1$, and the exact flux is specified elsewhere. Table \ref{sinLepErr} shows the errors and iteration numbers.
\begin{table}[hbtp]
   \footnotesize
   \begin{center}
      \begin{tabular}{|c|c c c c|c c c c|}
         \hline
         & \multicolumn{4}{|c|}{Current scheme} & \multicolumn{4}{|c|}{Previous scheme} \\
         \hline
         $h$ & 1/10 & 1/20 & 1/40 & 1/80 & 1/10 & 1/20 & 1/40 & 1/80 \\
         \hline
         \multicolumn{9}{|c|}{Hexahedral grids} \\
         \hline
         $\varepsilon_2^c$ & 2.39e-3 & 5.98e-4 & 1.49e-4 & 3.73e-5 & 5.28e-3 & 1.04e-3 & 2.17e-4 & 5.34e-5 \\
         $\varepsilon_\text{max}^c$ & 6.66e-3 & 1.69e-3 & 4.25e-4 & 1.06e-4 & 3.22e-2 & 8.07e-3 & 2.94e-3 & 7.78e-4 \\
         $\varepsilon_2^{\mathbf u}$ & 4.21e-3 & 1.05e-3 & 2.62e-4 & 6.56e-5 & 3.14e-2 & 1.08e-2 & 3.56e-3 & 1.46e-3 \\
         $\varepsilon_\text{max}^{\mathbf u}$ & 1.28e-2 & 3.20e-3 & 7.99e-4 & 1.99e-4 & 2.79e-1 & 1.60e-1 & 8.50e-2 & 4.13e-2 \\
         $n_6$ & 1 & 1 & 1 & 1 & 2 & 4 & 4 & 4 \\
         $n_9$ & 1 & 1 & 1 & 1 & 4 & 6 & 7 & 7 \\
         \hline
         \multicolumn{9}{|c|}{Prismatic grids} \\
         \hline
         $\varepsilon_2^c$ & 1.43e-3 & 3.58e-4 & 8.92e-5 & 2.12e-5 & 3.53e-3 & 8.05e-4 & 1.51e-4 & 3.13e-5 \\
         $\varepsilon_\text{max}^c$ & 5.53e-3 & 1.78e-3 & 4.18e-4 & 1.07e-4 & 2.36e-2 & 8.21e-3 & 2.05e-3 & 5.19e-4 \\
         $\varepsilon_2^{\mathbf u}$ & 4.90e-3 & 1.26e-3 & 3.96e-4 & 1.92e-4 & 2.80e-2 & 1.07e-2 & 3.63e-3 & 1.31e-3 \\
         $\varepsilon_\text{max}^{\mathbf u}$ & 5.77e-2 & 2.03e-2 & 1.22e-2 & 8.82e-3 & 3.79e-1 & 2.06e-1 & 1.02e-1 & 5.36e-2 \\
         $n_6$ & 6 & 8 & 11 & 17 & 6 & 8 & 11 & 16 \\
         $n_9$ & 10 & 15 & 22 & 36 & 10 & 14 & 21 & 35 \\
         \hline
         \multicolumn{9}{|c|}{Tetrahedral grids} \\
         \hline
         $\varepsilon_2^c$ & 1.03e-3 & 1.74e-4 & 3.45e-5 & 7.08e-6 & 1.02e-3 & 1.74e-4 & 3.48e-5 & 7.10e-6 \\
         $\varepsilon_\text{max}^c$ & 1.21e-2 & 1.50e-3 & 3.92e-4 & 9.44e-5 & 1.23e-2 & 1.50e-3 & 4.00e-4 & 1.02e-4 \\
         $\varepsilon_2^{\mathbf u}$ & 2.49e-2 & 8.98e-3 & 4.19e-3 & 1.99e-3 & 2.31e-2 & 8.56e-3 & 4.09e-3 & 1.96e-3 \\
         $\varepsilon_\text{max}^{\mathbf u}$ & 2.02e-1 & 1.49e-1 & 6.59e-2 & 4.21e-2 & 2.03e-1 & 1.01e-1 & 4.62e-2 & 2.53e-2 \\
         $n_6$ & 54 & 88 & 147 & 231 & 56 & 89 & 154 & 239 \\
         $n_9$ & 91 & 164 & 313 & 571 & 94 & 166 & 323 & 579 \\
         \hline
      \end{tabular}
   \end{center}
   \caption{Errors and iteration numbers in Example \ref{ex:sinuslepotier}.}
   \label{sinLepErr}
\end{table}

In this example the scheme reduces to a linear one on structured grids, therefore convergence is achieved in a single iteration. This was not the case with scheme \cite{Vid13}, owing to the treatment of Dirichlet boundary conditions. The change also results in smaller errors on structured and prismatic grids. This is only possible because the anisotropy axes are aligned with the grid, which was not the case in the previous example.

The present example was also solved on Kershaw meshes \cite{Ker81}. The errors and iteration numbers are shown in Table \ref{lepRobKerErr}. No more than one candidate set extension was necessary to find the cells and boundary conditions that give non-negative coefficients $\boldsymbol\alpha$ in (\ref{uapprox}), in comparison to \cite{Vid13} where up to 11 extension rounds were necessary to find convex combinations on the same meshes. As explained in \cite{Vid13}, to find a convex combination for a mesh node that does not belong to the convex hull of nearby collocation points, a large number of extensions is necessary. Kershaw grids contain such nodes. On the other hand, the new scheme does not use interpolated point values so it does not suffer from this problem.
\begin{table}[hbtp]
   \footnotesize
   \begin{center}
      \begin{tabular}{|c|c c c c c|}
         \hline
         & \multicolumn{5}{|c|}{Current scheme} \\
         \hline
         $h$ & 1/8 & 1/16 & 1/32 & 1/64 & 1/128 \\
         \hline
         $\varepsilon_2^c$ & 2.38e-2 & 2.08e-2 & 1.44e-2 & 6.62e-3 & 2.21e-3 \\
         $\varepsilon_\text{max}^c$ & 1.21e-1 & 9.32e-2 & 6.06e-2 & 2.85e-2 & 9.21e-3 \\
         $\varepsilon_2^{\mathbf u}$ & 3.67e-1 & 2.55e-1 & 1.52e-1 & 7.56e-2 & 2.95e-2 \\
         $\varepsilon_\text{max}^{\mathbf u}$ & 2.52 & 1.92 & 9.04e-1 & 5.58e-1 & 3.21e-1 \\
         $n_6$ & 59 & 118 & 203 & 381 & 635 \\
         $n_9$ & 91 & 186 & 331 & 670 & 1239 \\
         \hline
         & \multicolumn{5}{|c|}{Previous scheme} \\
         \hline
         $\varepsilon_2^c$ & 2.11e-1 & 2.07e-2 & 1.44e-2 & 6.62e-3 & 2.21e-3 \\
         $\varepsilon_\text{max}^c$ & 7.91e-1 & 9.38e-2 & 6.06e-2 & 2.84e-2 & 9.21e-3 \\
         $\varepsilon_2^{\mathbf u}$ & 5.51e-1 & 2.41e-1 & 1.52e-1 & 7.51e-2 & 2.93e-2 \\
         $\varepsilon_\text{max}^{\mathbf u}$ & 3.90 & 1.42 & 8.37e-1 & 5.34e-1 & 3.13e-1 \\
         $n_6$ & 63 & 118 & 206 & 384 & 636 \\
         $n_9$ & 97 & 199 & 335 & 674 & 1241 \\
         \hline
      \end{tabular}
   \end{center}
\caption{Errors and iteration numbers on Kershaw grids in Example \ref{ex:sinuslepotier}.}
\label{lepRobKerErr}
\end{table}
\end{example}

\begin{example}
The diffusion tensor is
\begin{equation}
\mathbb D=\left\{\begin{array}{c c }\mathbb D_1 & \text{if } x<0.1 \\ \mathbb D_2 & \text{if } x<0.95-0.75y \\ \mathbb D_3 & \text{otherwise,} \end{array}\right.
\end{equation} 
\begin{equation}
\mathbb D_1 = \left[\begin{array}{c c c}3&1&0\\1&3&0\\0&0&1\end{array}\right], \quad
\mathbb D_2 = \left[\begin{array}{c c c}10&3&0\\3&1&0\\0&0&1\end{array}\right], \quad
\mathbb D_3 = \left[\begin{array}{c c c}1.2&1&0\\1&2&0\\0&0&1\end{array}\right].
\end{equation} 
The exact concentration is a piecewise linear function
\begin{equation}
C=\left\{\begin{array}{c c}1+6x+4y & \text{if } x<0.1 \\
1.5+7x+4y & \text{if } x<0.95-0.75y \\ -2.3+5x+7y & \text{otherwise.}\end{array}\right.
\end{equation}
The exact concentration is set at $z=0$ and $z=1$, the exact flux is specified at $y=0$ and $y=1$, while the Robin condition is set elsewhere with $g_\text R|_{x=0}\equiv0$, $g_\text R|_{x=1}\equiv11$, and with $\Psi$ chosen accordingly. The problem was solved on the unit cube using tetrahedral, prismatic and hexahedral grids of various sizes. The obtained concentration was exact in all cases.
\end{example}

\begin{example}\label{noFlow}
This example was taken from \cite{Dan09}. The domain is a unit cube with two holes $[0,1]^3\backslash S_1\backslash S_2$, $S_1=[3/11,4/11]\times[5/11,6/11]\times[0,1]$, $S_2=[7/11,8/11]\times[5/11,6/11]\times[0,1]$. The diffusion tensor is
\begin{equation}
\mathbb D=R_z(-\theta)\diag (1,10^{-3},1) R_z(\theta),\label{rotatedTensor}
\end{equation}
where $R_z(\theta)$ is the matrix of rotation by angle $\theta=67.5\,^{\circ}$ around the $z$ axis. Concentration 0 is set in $S_1$, concentration 1 is set in $S_2$, and the no-flow condition is specified at the remaining boundary.

The example was solved on a range of hexahedral grids. The minimal and maximal concentrations and the iteration numbers are shown in Table \ref{noFlowTable}. This example demonstrates that the scheme preserves positivity. The maximum principle violation is somewhat lesser than in \cite{Dan09}, owing to the implicit treatment of Neumann boundary conditions.
\begin{table}[hbtp]
   \footnotesize
   \begin{center}
      \begin{tabular}{|c|c c c c|c c c c|}
         \hline
         & \multicolumn{4}{|c|}{Current scheme} & \multicolumn{4}{|c|}{Previous scheme} \\
         \hline
         $h$ & 1/11 & 1/22 & 1/44 & 1/88 & 1/11 & 1/22 & 1/44 & 1/88 \\
         \hline
         $C_\text{min}$ & 1.50e-2 & 3.55e-3 & 5.66e-4 & 5.36e-5 & 1.47e-2 & 3.51e-3 & 5.61e-4 & 5.33e-5 \\
         $C_\text{max}$ & 1.47 & 1.35 & 1.08 & 1.02 & 1.56 & 1.38 & 1.08 & 1.02 \\
         $n_6$ & 119 & 360 & 816 & 1063 & 129 & 346 & 813 & 1063 \\
         $n_9$ & 230 & 788 & 1998 & 3136 & 240 & 777 & 2001 & 3138 \\
         \hline
      \end{tabular}
   \end{center}
\caption{Concentration span and iteration numbers in Example \ref{noFlow}.}
\label{noFlowTable}
\end{table}
\end{example}

\begin{example}\label{noFlowDiscontinuous}
Example \ref{noFlow} was modified to demonstrate that positivity is preserved in the presence of discontinuities. The domain is $[0,1]^3\backslash S_1\backslash S_2$, $S_1=[0.2,0.4]\times[0.4,0.6]\times[0,1]$, $S_2=[0.6,0.8]\times[0.4,0.6]\times[0,1]$. The diffusion tensor is given by (\ref{rotatedTensor}),
where $\theta=22.5\,^{\circ}$ if $(x<0.5\land y<0.5)\lor(x>0.5\land y>0.5)$ and $\theta=67.5\,^{\circ}$ otherwise. The boundary conditions are the same as in the previous example.

The example was solved on a range of hexahedral grids. The minimum and maximum concentration values, as well as the iteration numbers, are given in Table \ref{noFlowDiscontinuousTable}. Positivity is preserved but, unlike the previous example, it is not evident that the maximum principle violation reduces as the grid is refined.
\begin{table}[hbtp]
   \footnotesize
   \begin{center}
      \begin{tabular}{|c|c c c c|c c c c|}
         \hline
         & \multicolumn{4}{|c|}{Current scheme} & \multicolumn{4}{|c|}{Previous scheme} \\
         \hline
         $h$ & 1/10 & 1/20 & 1/40 & 1/80 & 1/10 & 1/20 & 1/40 & 1/80 \\
         \hline
         $C_\text{min}$ & 2.57e-3 & 1.06e-3 & 1.18e-4 & 7.47e-6 & 1.20e-3 & 1.05e-3 & 1.15e-4 & 7.23e-6 \\
         $C_\text{max}$ & 1.17 & 1.53 & 1.72 & 1.33 & 1.06 & 1.49 & 1.731 & 1.34 \\
         $n_6$ & 76 & 188 & 461 & 861 & 103 & 213 & 484 & 877 \\
         $n_9$ & 140 & 389 & 1090 & 2457 & 185 & 432 & 1133 & 2475 \\
         \hline
      \end{tabular}
      \end{center}
      \caption{Concentration span and iteration numbers in Example \ref{noFlowDiscontinuous}.}
      \label{noFlowDiscontinuousTable}
\end{table}
\end{example}

\begin{example}\label{ex:discontiniousdanilovtransient}

We modified Example \ref{ex:discontiniousdanilov} to test the behavior of the new scheme in the transient case. The exact solution is
\begin{equation}
C=e^t\left\{\begin{array}{c}1-2y^2+4xy+2y+6x \quad \text{if } x<0.5,\\3.5-2y^2+2xy+x+3y \quad \text{otherwise.}\end{array}\right.
\end{equation}
The initial solution at $t=0$ and the source term are chosen accordingly. The boundary conditions and the diffusion tensor are as in Example \ref{ex:discontiniousdanilov}. 

The problem was solved on the same series of tetrahedral grids as in Example \ref{ex:discontiniousdanilov} using one, two, and three iterations per time step $\Delta t=1.6h^2$. Such a time step was chosen because it generates an error of the same order of magnitude as the spatial discretization error. For comparison, on the coarser grids the problem was also solved using $\Delta t=10^{-5}$ and as many iterations as necessary to reach $\varepsilon = 10^{-9}$. The errors at $t=2$ given in Table \ref{disDanTranErr} show that a single iteration is sufficient to maintain second order accuracy, although the maximal flux error is significantly reduced if two or three iterations are performed.
\begin{table}[hbtp]
   \footnotesize
   \begin{center}
      \begin{tabular}{|c|c c c c|c c c c|}
         \hline
         $h$ & 1/10 & 1/20 & 1/40 & 1/80 & 1/10 & 1/20 & 1/40 & 1/80 \\
         \hline
         & \multicolumn{4}{|c|}{$\Delta t=1.6h^2$, 1 iteration} & \multicolumn{4}{|c|}{$\Delta t=1.6h^2$, 2 iterations} \\
         \hline
         $\varepsilon_2^c$ & 2.88e-3 & 6.48e-4 & 1.36e-4 & 3.23e-5 & 2.34e-3 & 4.77e-4 & 8.87e-5 & 2.20e-5 \\
         $\varepsilon_\text{max}^c$ & 2.88e-2 & 7.70e-3 & 1.93e-3 & 5.73e-4 & 2.19e-3 & 6.39e-3 & 1.71e-3 & 5.39e-4 \\
         $\varepsilon_2^{\mathbf u}$ & 7.71e-3 & 2.98e-3 & 1.32e-3 & 6.02e-4 & 5.59e-3 & 2.42e-3 & 1.17e-3 & 5.64e-4 \\
         $\varepsilon_\text{max}^{\mathbf u}$ & 7.46e-2 & 6.19e-2 & 3.23e-2 & 1.93e-02 & 5.27e-2 & 3.05e-2 & 9.84e-3 & 8.44e-3 \\
         \hline
         & \multicolumn{4}{|c|}{$\Delta t=1.6h^2$, 3 iterations} & \multicolumn{4}{|c|}{$\Delta t=10^{-5}$, $\varepsilon = 10^{-9}$} \\
         \hline
         $\varepsilon_2^c$ & 2.41e-3 & 5.34e-4 & 1.07e-4 & 2.60e-5 & 1.65e-3 & 3.85e-4 &  &  \\
         $\varepsilon_\text{max}^c$ & 2.09e-2 & 6.29e-3 & 1.71e-3 & 5.40e-4 & 1.72e-2 & 5.47e-3 &  &  \\
         $\varepsilon_2^{\mathbf u}$ & 5.23e-3 & 2.37e-3 & 1.15e-3 & 5.61e-4 & 5.12e-3 & 2.36e-3 &  &  \\
         $\varepsilon_\text{max}^{\mathbf u}$ & 3.72e-2 & 1.44e-2 & 8.24e-3 & 5.60e-3 & 3.60e-2 & 1.45e-2 &  &  \\
         \hline
      \end{tabular}
      \label{disDanTranErr}
   \end{center}
\caption{Concentration errors in Example \ref{ex:discontiniousdanilovtransient}.}
\end{table}

\end{example}

\section{Discussion and conclusion}

We have demonstrated that there is little difference in accuracy and convergence rate between the current scheme and \cite{Vid13}. They both preserve solution positivity. Since the current scheme is much simpler, it should be used instead of \cite{Vid13}.

To construct conical combinations (\ref{uapprox}), a more sophisticated search strategy could be devised, such as the one presented in \cite{Vid13}. However, the need to do so is lesser than in the case of convex combinations because the computational effort is smaller. There are two reasons for this. The first reason is that we have not encountered a case where the cells and boundary conditions used to construct system (\ref{system}) were more than three cells away. For comparison, up to 11 candidate set extension rounds were necessary to represent Kershaw grid node values as convex combinations in \cite{Vid13}. Second, three vectors suffice to construct a conical combination, while in general four points are necessary for a convex combination --- so the systems are smaller and fewer in the case of conical combinations, even for the same number of extensions.

In practice, we have never come across a situation in which the proposed algorithm did not yield a conical combination. However, this may happen in the case presented in \cite{Vid13}, where the diffusion tensor varies in a sharp domain corner. In such a case we suggest modifying the mesh. Situations where a discontinuity meets the boundary, which required the introduction of auxiliary variables in the earlier versions of the scheme, do not present a difficulty in the current version.

Maximum principle violations are due to the way the one-side fluxes are combined in \S \ref{sec:scheme} and can be severe, as shown in Example \ref{noFlowDiscontinuous}. Nevertheless, this is not related to the technique of using piecewise linear transformation in linear flux reconstruction. The same technique can be used in combination with a variety of other finite volume schemes, such as \cite{Dro11,She11}.

It was demonstrated that in the transient case a single iteration per time step is enough to maintain the second order spatial accuracy with Euler implicit scheme, provided that sufficiently small time steps and the proposed initial guesses are used.

\vspace{1ex}
{\bf Acknowledgments.} The authors wish to thank the Ministry of Education, Science and Technological Development of the Republic of Serbia for the financial support provided through Technology Development Project TR37014.

\bibliographystyle{elsarticle-num}
\bibliography{biblioT}

\end{document}